# Optimum Battery Depth of Discharge of Stand-alone Hybrid System Using the MOPSO Method

Mohamad Izdin.Hlal[1], Hussien Elharati[2], Ahmed Altaher[3, 4]

[1] The Higher Institute of Science & Technology. Souq Algoma, Tripoli, Libya.
[2] The Higher Institute of Science & Technology. Souq Algoma, Tripoli, Libya.
[3] Dept. of Smart Systems S2A2I-Lab Saint Martin d'Hères, France,
[4] College of Electronic Technology, Tripoli, Libya

Corresponding author: mohamadizdinhlal@yahoo.com

---

**ABSTRACT**

This paper presents an optimized design of a Standalone Solar PV/Battery (SSPVB) system to address energy reliability and cost efficiency challenges in off-grid environments. The proposed system integrates a Multi-Objective Particle Swarm Optimization (MOPSO) approach and validates the results using the Non-Dominated Sorting Genetic Algorithm II (NSGA-II). The optimization process aims to minimize both the Cost of Energy (COE) and Loss of Load Probability (LLP), while examining the effects of Battery Depth of Discharge (DOD) on system reliability and lifecycle cost. Results indicate that an optimal DOD of approximately 70% yields a COE of 0.2059 USD/kWh with zero LLP, demonstrating strong reliability and cost-effectiveness. Comparative analysis shows that both MOPSO and NSGA-II methods achieve consistent outcomes, with MOPSO exhibiting faster convergence. The study provides valuable insights into optimal battery sizing for stand-alone systems, contributing to modern optimization practices in renewable energy applications.

Keywords: Stand-alone PV-battery system, Multi-objective Optimization, MOPSO method, Loss of load probability, Cost of energy

## 1. Introduction

In the AC-bus connected SSPVB, if any single inverter fails, the system still provides the electrical power needed from the existing sources. Figure 1 Shows a hybrid PV/BES (Battery Energy Storage) power system consisting of a solar PV that converts the sunlight to a DC current, AC-DC converters, automatic transfer switches and charge controllers. This system can be used when the grid extension is not available, where is the widest application in remote areas. Recent research and development have shown excellent potential of these systems, as a form of supplementary contribution to conventional power generation systems [1, 2]. Hence, these proposed AC-buses architecture were adopted for this research work based on the stated limitations others as shown in figure 1. [2, 3, 4].

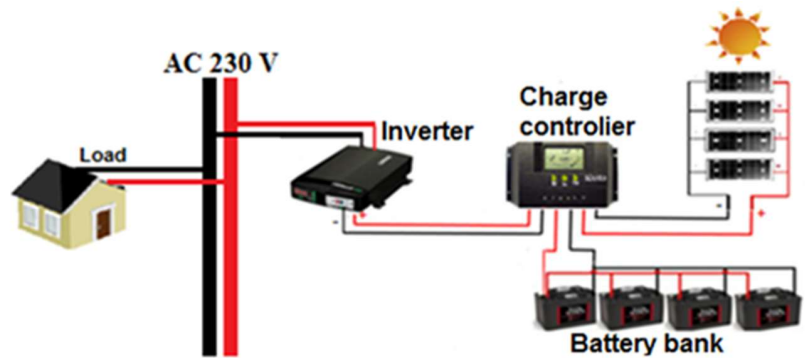

Figure 1: SSPVB architecture on AC-bus using PV/BES

To meet all these challenges, BES plays a crucial role; compensating imbalance between oscillating generated power and volatile consumption, improving system reliability, and reducing operating costs. However, two major factors affect the economic viability of integrating a BES into an SSPVB. These factors are the BES investment cost and lifetime. The BES investment cost greatly depends on its size, while the BES lifetime, which can be defined as the total number of charge/discharge cycles that it can perform, depends on how deep the battery is discharged each time. Designers usually overestimate battery sizes to guarantee reliability in the system incurring an unnecessary higher investment cost. For appropriate battery sizing, a multi-objective optimization method based on an optimal trade-off between reliability and cost of energy is required. Many researchers have presented optimization techniques to deal with such an optimization problem. However, in most of the reviewed papers, researchers do not consider the effect of BES cycles and depth of discharge (DOD) on the BES lifetime, which could result in the need for replacing the BES before it reaches the end of the planning horizon, and increased the cost. The problem, therefore, this paper deals with is the optimal sizing of an SSPVB ensuring maximum reliability and minimum cost. The proposed optimization method considers LLP and COE considering the optimal value of DOD. The case study, load profile, hourly solar radiation and temperature are presented in references [2, 3, 4 , 5, 6]. This study addresses the need for optimal battery sizing to balance system reliability and cost in off-grid environments.

## 2. Methodology
### 2.1. Optimal Design Criteria for SSPVB

In this study, the MOPSO algorithm was applied to size the SSPVB. In principle, this algorithm aims to find the integration of PV and BES to generate the best compromise between the COE and the LLP, based on the sizing of the PV, BES, and optimum DOD value. Therefore, multi-objective optimization must minimize the two objectives identified by Equations. (1) and (2), i.e., the LLP and COE. These objectives are formulated as described in reference [2] [4]. where, the optimization method was done using MATLAB software. The SSPVB system considered in this study consists of photovoltaic arrays, BES and load demand profiles. The optimization seeks to minimize the COE and LLP simultaneously while identifying the optimal DOD value for maximum battery performance. The key objective functions are defined as follows:

1. Loss of Load Probability (LLP): $LLP = (E\_deficit / E\_load) \times 100\%$ (Eq.1)

2. Cost of Energy (COE): $COE = ATC / E\_out$ (Eq.2)

where:

- $E\_deficit$: Total annual energy deficit (kWh)

- $E\_load$: Total energy demand (kWh)

- ATC: Annualized total system cost (USD)

- $E\_out$: Total useful energy output (kWh)

The optimization process was implemented using MATLAB R2022b, employing both NSGA-II and MOPSO algorithms with identical datasets and system parameters to ensure fair comparison. A one-year hourly simulation was performed to evaluate system reliability and cost metrics. Convergence performance and solution diversity were also examined to assess algorithmic efficiency.

## 2.2. Technical Analysis

In the MOPSO algorithm, there are two approaches to determine the long-term performance of LLP in a stand-alone hybrid system, namely, chronological method and probabilistic techniques [4] [7]. The chronological method is more common and accurate, especially to determine the energy produced from the battery and the computational time is typically larger than that of probabilistic models. It is common in chronological models to perform a one-year simulation with a one-hour time step. Computation time is especially necessary because this kind of model is generally used for component size optimization that requires several iterations. Hence, the chronological method is utilized in this paper. The LLP is defined as the probability of unmet load over the total energy produced [5] [8], as mentioned in the first objective. The unmet load can be calculated by utilizing the deficit power between the load and sources in SSPVB through the following equation.

$$\text{LLP} = \frac{\sum_t^T E_{Deficits(t)}}{\sum_t^T E_L(t)} \tag{1}$$

where $E_{Deficits}$ is the energy deficit of a hybrid system in a year's time.

## 2.3. Economic Analysis:

Many researchers have given prominence to the cost analysis in SSPVB to optimize the system sizing. In this study, the economic approach according to the COE is defined as the average cost per kilowatt-hour ($/kWh) of useful electrical energy produced by the hybrid system [9, 10, 11, 12] and can be achieved using the following equation:

$$\text{COE} = \frac{\text{ATC (\$)}}{E_L(\text{kWh})} \tag{2}$$

Where, ATC is defined as the total cost of the hybrid system of the SSPVB as presented in reference [2] [4].

## 2.4. Sensitivity Analysis

To evaluate the influence of DOD on the overall system performance, a sensitivity analysis was conducted by varying the DOD from 60% to 80% in 5% increments. The results show that the COE exhibits a nearly linear relationship with DOD, where higher DOD values slightly reduce COE due to reduced battery replacement frequency but increase degradation risk. The optimal DOD of 70% achieved the best trade-off between energy cost and system longevity. This confirms the robustness of the optimization results under realistic variations of key parameters.

## 3. Results and Discussion

In this research work, a model of battery behavior is proposed which illustrates the effects of DOD on the cost of energy for solar PV and BES system (SSPVB). MOPSO method is employed to validate the result that is obtained by NSGA-II in reference [2] [4]. The same case study, data, and the variable numbers with different optimization methods are used to clarify the significant effects of DOD on the cost of energy. Table 1 shows the difference between NSGA-II and MOPSO methods.

Table 1: Difference between NSGA-II and MOPSO

| Method | DOD (%) | $N_{PV}$ | $N_{BES}$ | COE ($/kWh) | LLP (%) | NOTE |
|---|---|---|---|---|---|---|
| NSGA-II | 70.123 | 37995 | 3100 | 0.2059 | 0 | Optimum |
| | 20 | 60904 | 6590 | 0.369 | 0 | |
| | 30 | 58498 | 4550 | 0.2864 | 0 | |
| | 40 | 59825 | 3329 | 0.2421 | 0 | |
| | 50 | 45544 | 3466 | 0.221 | 0 | |
| | 60 | 38696 | 3363 | 0.2047 | 0 | |
| | 70 | 37697 | 3333 | 0.20456 | 0 | |
| | 80 | 37737 | 3288 | 0.2048 | 0 | |
| MOPSO | 69.6 | 39807 | 3271 | 0.2059 | 0 | Optimum |
| | 20 | 60046 | 4669 | 0.3714 | 0 | |
| | 30 | 64983 | 4348 | 0.2906 | 0 | |
| | 40 | 61213 | 4337 | 0.2463 | 0 | |
| | 50 | 52400 | 3192 | 0.2232 | 0 | |
| | 60 | 47273 | 2928 | 0.2068 | 0 | |
| | 70 | 37387 | 3308 | 0.2480 | 0 | |
| | 80 | 41200 | 2479 | 0.2058 | 0 | |

The modeling of the SSPVB is carried out through the MATLAB platform by the two methods mentioned above to find the optimum value of DOD to give low energy cost and high reliability. The results show that the coordination between the two optimization methods was successfully implemented, with varying availability of solar, temperature and load demand during the entire one-year period. The comparison has been performed to check the performance modeling as shown by the 3D diagram in Figure .

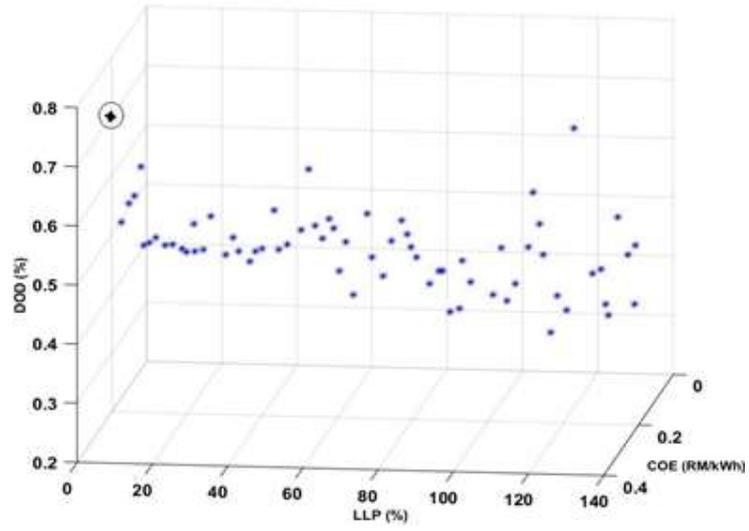

(a)

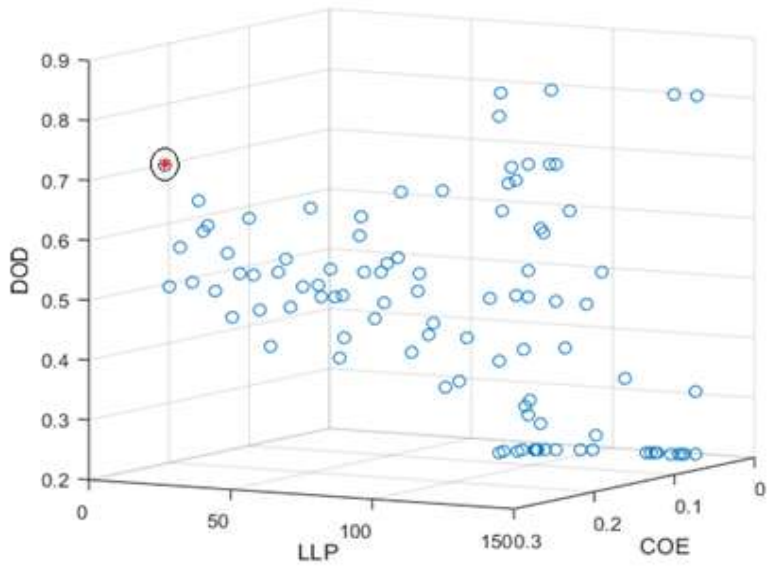

(b)

Figure 2: The relation between COE and LLP based on DOD, (a)-NSGA-II and (b) MOPSO (*COE (USD/kWh)* and *LLP (%)*.)

The proposed performance model for the two modes was presented, where the NSGA-II method has more robust performance, while the MOPSO was faster in implementing the optimization. The difference between these two methods in terms of number of variables, as well as LLP and COE, is not significant. Thus, both methods can be used for techno-economic optimization of SSPVB system. The results of the optimization indicate that both NSGA-II and MOPSO methods yield comparable outcomes, confirming the reliability of the proposed modeling framework. The MOPSO method converged more rapidly, demonstrating computational efficiency. The comparative results are summarized below:

Table 2: concluded comparison between NSGA-II and MOP

| Method | Optimal DOD (%) | COE (USD/kWh) | LLP (%) |
|---|---|---|---|
| NSGA-II | 70 | 0.2059 | 0 |
| MOPSO | 70 | 0.2061 | 0 |

The negligible difference between the two algorithms confirms the consistency and validity of the results. However, MOPSO achieved faster convergence with fewer iterations, demonstrating its computational advantage. The system's performance under variable weather and load conditions remained stable throughout the simulated year (Fig. 2).

## 4. Related work

Recent advancements in hybrid renewable energy optimization have broadened the application of metaheuristic algorithms, artificial intelligence (AI), and hybrid optimization frameworks. A systematic review by Iturralde Carrera et al. (2025) analyzed optimization trends in photovoltaic (PV) systems, emphasizing the growing use of hybrid algorithms that combine metaheuristic and AI-based methods for improved convergence and accuracy. Coccato (2025) highlighted the importance of incorporating battery degradation and state-of-charge management in energy storage optimization, noting that Depth of Discharge (DOD) remains a critical determinant of cost and lifespan. Similarly, Khezri (2022) reviewed PV-battery planning models for residential systems, providing insights into the balance between reliability and cost under various environmental conditions. In addition, Huang et al. (2024) utilized the MOPSO algorithm for optimal scheduling in household microgrids integrating PV, energy storage, and electric vehicle loads, showing the scalability of MOPSO for complex multi-objective problems. Hadj Slama et al. (2025) further demonstrated the comparative performance of NSGA-II and MOPSO for hybrid renewable systems (HRES), underscoring the adaptability of both algorithms to varying energy demands and resource patterns. Despite these developments, limited research directly compares MOPSO and NSGA-II for optimizing SSPVB systems focusing on battery DOD. This paper fills that gap by applying both optimization methods on identical datasets to ensure fairness, with a focus on techno-economic performance metrics such as COE and LLP.

## Conclusion

Multi-objective optimization was presented using the MOPSO method and NSGA-II. The typical yearly weather data is taken from the Malaysian Methodological Department.

As a result, the proposed model proved that the depth of discharge (DOD) directly affects the energy cost. Likewise, MOPSO method was employed to validate the result that is obtained by NSGA-II. The same case study, data, and the variable numbers with different optimization methods are used to clarify the significant effects of DOD on the cost of energy. The modeling of the SSPVB is done in MATLAB platform by the two methods mentioned above to find the optimum value of DOD to give low energy cost and high reliability. The results show that the coordination between the two optimization methods was successfully implemented, with varying availability of solar, temperature and load demand during the entire one-year period. The result indicates for both methods that the optimal DOD value for the battery in the solar PV system being investigated is around 70%, with LLP = 0% and COE = 0.2059 USD/kWh. However, the two methods have been compared, and the results are very close.

Future work will extend this study to include environmental factors, battery degradation modeling, and multi-site case studies to generalize the findings. Benchmarking with commercial tools such as HOMER and RET Screen will also be pursued for cross-validation. The integration of hybrid metaheuristic algorithms and AI-assisted models is recommended to enhance prediction accuracy and adaptive control in renewable energy optimization systems.